# Tuning Electroluminescence from Functionalized SWCNT Networks further into the Near-Infrared

*Nicolas F. Zorn, Simon Settele, Finn L. Sebastian, Sebastian Lindenthal and Jana Zaumseil\**

Institute for Physical Chemistry, Universität Heidelberg, D-69120 Heidelberg, Germany

Corresponding Author
*E-mail: zaumseil@uni-heidelberg.de

ABSTRACT

Near-infrared electroluminescence from carbon-based emitters, especially in the second biological window (NIR-II) or at telecommunication wavelengths, is difficult to achieve. Single-walled carbon nanotubes (SWCNTs) have been proposed as a possible solution due to their tunable and narrowband emission in the near-infrared and high charge carrier mobilities. Furthermore, the covalent functionalization of SWCNTs with a controlled number of luminescent *sp³* defects leads to even more red-shifted photoluminescence with enhanced quantum yields. Here, we demonstrate that by tailoring the binding configuration of the introduced *sp³* defects and hence tuning their optical trap depth we can generate emission from polymer-sorted (6,5) and (7,5) nanotubes that is mainly occurring in the telecommunication O-band (1260-1360 nm). Networks of these functionalized nanotubes are integrated in ambipolar, light-emitting field-effect transistors to yield the corresponding narrowband near-infrared electroluminescence. Further investigation of the current and carrier density-dependent electro- and photoluminescence spectra enable insights into the impact of different *sp³* defects on charge transport in networks of functionalized SWCNTs.

**Introduction**

The near-infrared wavelength range is highly interesting for applications in biological imaging (first and second biological window, NIR-I and NIR-II)[1, 2] as well as telecommunication (O-band at 1260-1360 nm and C-band at 1550 nm). For many of the corresponding light-emitting devices it would be advantageous to move away from conventional inorganic semiconductors and toward solution-processable and flexible materials. However, in stark contrast to the visible range, narrowband near-infrared electroluminescence beyond 800 nm, and even more so beyond 1000 nm, is very difficult to achieve with organic emitters. While a number of organic semiconductors with small bandgaps exists (e.g., donor-acceptor polymers[3]), which even show high charge carrier mobilities, they all suffer from very low (below 1 %) photoluminescence quantum yields (PLQY),[4-6] often rationalized with the gap law.[7] Consequently, only very few examples of near-infrared organic light-emitting diodes or light-emitting transistors have been demonstrated at all. More efficient near-infrared emitters always contain heavier elements such as inorganic quantum dots ($Ag_2S$, InAs)[8, 9] or rare-earth complexes.[10]

Semiconducting single-walled carbon nanotubes (SWCNTs) are a potential alternative as they combine solution-processability, very high hole and electron mobilities and narrowband emission in the near-infrared, which can be tuned from 900 nm to over 1600 nm depending on the diameter and (n,m) species (i.e., chirality) of the nanotubes.[11] Several methods for the separation of single nanotube species have been developed and optimized over the past decade to obtain large amounts of pure material.[12] Thus, it is now possible to create optoelectronic devices, including photovoltaic cells,[12] electrochromic cells,[13] light-emitting diodes[14] and light-emitting transistors[15, 16] based on dense networks of a single (n,m) types of nanotubes with narrow and well-controlled absorption and emission bands. However, the more abundant and easiest to purify nanotube species are those with emission in the range of 950 to 1100 nm, such as the (6,5) and (7,5) nanotubes.

Unfortunately, the PLQY of SWCNTs in the near-infrared still only reaches few percent in dispersion[17] and is even lower in dense networks. In addition to the gap law, dark states, and Auger-quenching by charge carriers,[18] the fast diffusion of excitons within nanotubes toward potential quenching sites (e.g., the ends of the nanotubes)[19] is assumed to be the main cause for the observed low emission efficiencies. The controlled introduction of luminescent *sp*³ defects in SWCNTs[20, 21] has been shown to be an effective path to reduce undesired quenching, increase PLQY and shift the emission wavelength even further into the near-infrared. A small number of specific lattice defects along the nanotube create exciton trap states where previously mobile excitons are localized. They subsequently decay radiatively from the defect state at longer wavelengths than the mobile excitons (optical trap depths of 150-250 meV).[22] Due to this efficient localization, the excitons are unable to reach other non-radiative defects and hence the overall PLQY is increased especially for very short nanotubes.[23]

Several different methods of functionalization to create luminescent defects have been developed over the past decade that can be applied to either surfactant-stabilized nanotubes in water[21, 24] or polymer-wrapped nanotubes in organic solvents.[13, 25] Depending on the employed chemistry, different types of lattice defects are introduced that result in more or less red-shifted emission peaks. Here we will refer to the commonly observed defect emission with optical trap depths of 160 - 190 meV as $E_{11}*$ emission and to the further red-shifted emission as $E_{11}*^-$ (optical trap depths >200 meV).[26] Most studies on *sp*³ defects in SWCNTs are carried out on (6,5) nanotubes (diameter 0.76 nm, optical bandgap ~ 1.27 eV) due to their excellent availability as pure dispersions and their high reactivity to form *sp*³ defects. Maximum PLQYs of up to 4-5 % are typically reached in dispersions of polymer-wrapped (6,5) SWCNTs with on average 5 to 8 defects per micrometer.[25, 27, 28] However, for further red-shifted emission, SWCNTs with larger diameters are preferred, such as (7,5), (9,4) or (10,5) nanotubes. These have also been functionalized and in some cases even showed single-photon emission up to the C-band range at room temperature.[29, 30]

The integration of $sp^3$-functionalized SWCNTs in lateral light-emitting diodes and ambipolar field-effect transistors showing electroluminescence (EL) from defects has recently been demonstrated for individual nanotubes[31] as well as sparse[32] and dense networks.[33] The defect emission properties were maintained in these devices and charge transport was only slightly affected. However, those examples only involved SWCNTs with $E_{11}*$ defects and did not reach far into the near-infrared. Here we tailor the covalent functionalization of polymer-sorted (6,5) and (7,5) nanotubes to obtain $E_{11}*^-$ defect electroluminescence further in the near-infrared using light-emitting transistors with dense SWCNT networks. Emission in the telecommunication O-band is achieved without sacrificing the ambipolar charge transport properties required for high current densities. We further explore the correlation of optical trap depth and charge carrier trapping for both nanotube species as well as carrier density-dependent photoluminescence and electroluminescence spectra for additional device tunability.

**Results and Discussion**

To achieve further red-shifted photo- and electroluminescence from networks of polymer-wrapped (6,5) and (7,5) nanotubes, they were first selectively dispersed in toluene by shear-force mixing with PFO-BPy and PFO (see molecular structures in Figure 1), respectively, followed by centrifugation. Subsequently, they were functionalized with $E_{11}*^-$ defects using the established reaction with 2-iodoaniline in the presence of potassium *tert*-butoxide (KO$^t$Bu, see schematic in Figure 1 and Experimental Section for detailed description).[25] A portion of pristine nanotubes was retained for reference in both cases.

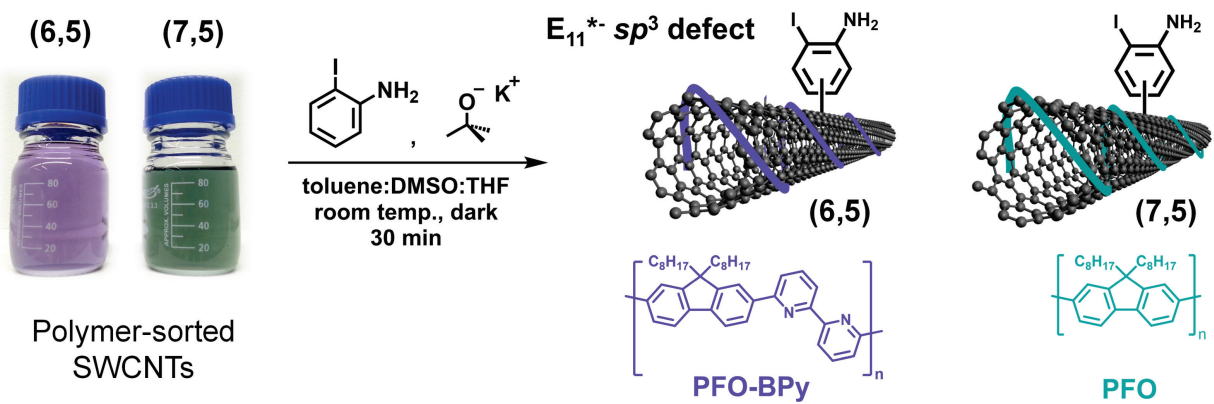

**Figure 1.** General reaction scheme for the covalent functionalization of PFO-BPy-wrapped (6,5) and PFO-wrapped (7,5) SWCNTs with 2-iodoaniline and potassium *tert*-butoxide (KO$^t$Bu) in organic solvents to introduce $E_{11}^{*-}$ defects. Note that bond formation between the defect moiety and the SWCNT might occur at different carbon atoms of the aromatic ring.

The successful introduction of $E_{11}^{*-}$ defects was confirmed by photoluminescence (PL) spectroscopy of the dispersions. Figure 2a shows the PL spectra of pristine and functionalized (6,5) SWCNTs in toluene normalized to the $E_{11}$ emission from mobile excitons at 1000 nm. Only one strongly red-shifted emission peak at 1251 nm resulting from $E_{11}^{*-}$ defects (optical trap depth 248 meV) is evident and no or only negligible emission from $E_{11}^{*}$ defects (expected at ~1170 nm) is observed. The amplitude averaged fluorescence lifetime of the defect emission was determined by time-correlated single photon counting (TCSPC) to be 362 ps, which is consistent with the deeper optical trap depth of $E_{11}^{*-}$ compared to $E_{11}^{*}$ defects. Likewise, a functionalized dispersion of (7,5) SWCNTs (see Figure 3a) exhibits an $E_{11}^{*-}$ defect emission peak at 1272 nm compared to the corresponding $E_{11}$ peak at 1048 nm. The optical trap depth for this defect in (7,5) nanotubes, however, is only 208 meV. A reduction of the optical trap depth with increasing nanotube diameter is a common observation for functionalized SWCNTs and was reported for this type of defect as well.[21, 25, 34] For both functionalized (6,5) and (7,5) nanotubes, the defect emission is close to or already within the O-band for telecommunication (1260-1360 nm). The corresponding absorption and Raman spectra before and after functionalization are shown in Figures S1 and S2 (Supporting Information).

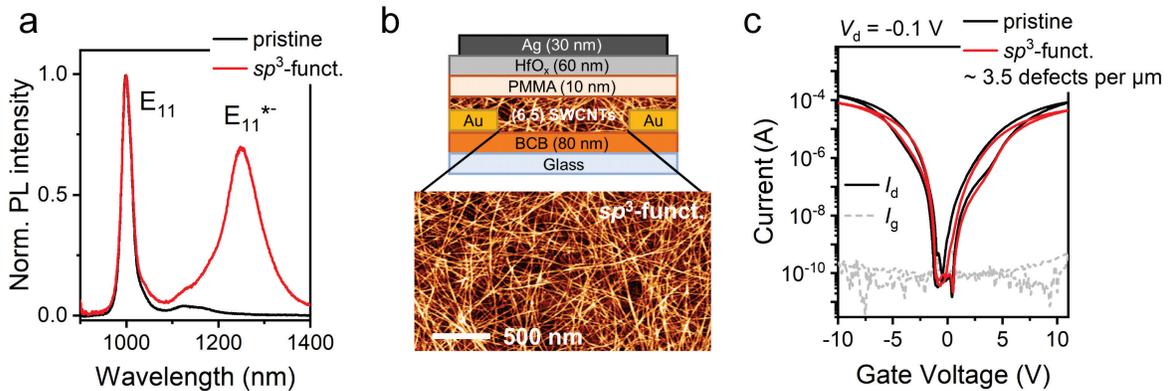

**Figure 2.** (a) Normalized PL spectra of a pristine (6,5) SWCNT dispersion and one after functionalization. Emission peaks corresponding to mobile $E_{11}$ excitons and defect-localized $E_{11}^{*-}$ excitons are labelled. (b) Schematic cross-section of a bottom-contact, top-gate (6,5) SWCNT network FET (layer thicknesses not to scale) and a representative atomic force micrograph of a dense network of *sp*$^3$-functionalized SWCNTs. The scale bar is 500 nm. (c) Ambipolar transfer characteristics of pristine and functionalized SWCNT network FETs in the linear regime ($V_d$ = -0.1 V). Solid lines represent drain currents $I_d$, grey dashed lines are gate leakage currents $I_g$.

The pristine and functionalized dispersions were used to create bottom-contact, top-gate field-effect transistors (FETs) with a bilayer gate dielectric of PMMA and HfO$_x$ as introduced previously[33] and schematically shown in Figure 2b. An additional surface passivation layer of BCB (a cross-linked benzocyclobutene-based polymer, see Experimental Section) on the glass substrate prevented undesired sideband emission and enabled nearly intrinsic PL and EL spectra.[35] The SWCNT network density (see inset in Figure 2b and Figure 3b) for all samples was chosen to be well above the percolation limit and above the mobility saturation for nanotube networks[36] to ensure comparability of the extracted mobility values. The obtained transfer characteristics of representative FETs in the linear regime (drain voltage $V_d$ = -0.1 V) are shown in Figures 2c and 3c for (6,5) and (7,5) SWCNT networks, respectively. All of them exhibit typical ambipolar charge transport with only little hysteresis and high on/off current ratios. The average values of the extracted hole and electron mobilities are summarized in Table S1 (Supporting Information) with the corresponding defect densities (3.5 and 7.6 µm$^{-1}$, respectively) that were determined based on the differential increase of the integrated Raman D/G$^+$ mode ratios as reported previously.[27] Note that the hole and electron mobilities of the (7,5) networks (~ 0.14 cm²V$^{-1}$s$^{-1}$) are much lower than those of the (6,5) nanotubes (~ 2-4 cm²V$^{-1}$s$^{-1}$), which is likely due to stronger bundling as visible in Figure 3b (average bundle thickness 5-8 nm *versus* 3-4 nm for (6,5) SWCNTs) and the presence of residual (7,6) nanotubes (see Figure S2) with a smaller bandgap that may act as trap sites.[37] The increased bundling of (7,5) nanotubes and hence the less efficient removal of adsorbed water from the network at moderate annealing temperatures may also explain the larger hysteresis for electron transport in these transistors (see Figure 3c).

Importantly, the reduction of the hole (from 3.9 to 2.2 cm²V$^{-1}$s$^{-1}$) and electron mobilities (from 2.1 to 1.1 cm²V$^{-1}$s$^{-1}$) for the (6,5) nanotube networks at a relatively low defect density indicates a similar if not stronger impact of $E_{11}^{*-}$ defects on charge transport within the nanotubes compared to the previously investigated $E_{11}^*$ defects.[33,38] The (7,5) nanotube networks show a similar behavior with a reduction of the mobility by half for slightly higher defect densities. The substantially deeper optical trap depth of $E_{11}^{*-}$ defects in (6,5) SWCNTs (268 meV) compared to both $E_{11}^*$ defects (160-190 meV) and $E_{11}^{*-}$ defects in (7,5) nanotubes (208 meV) may also reflect the trap depth for charge carriers and hence rationalize the stronger impact on charge carrier mobilities.

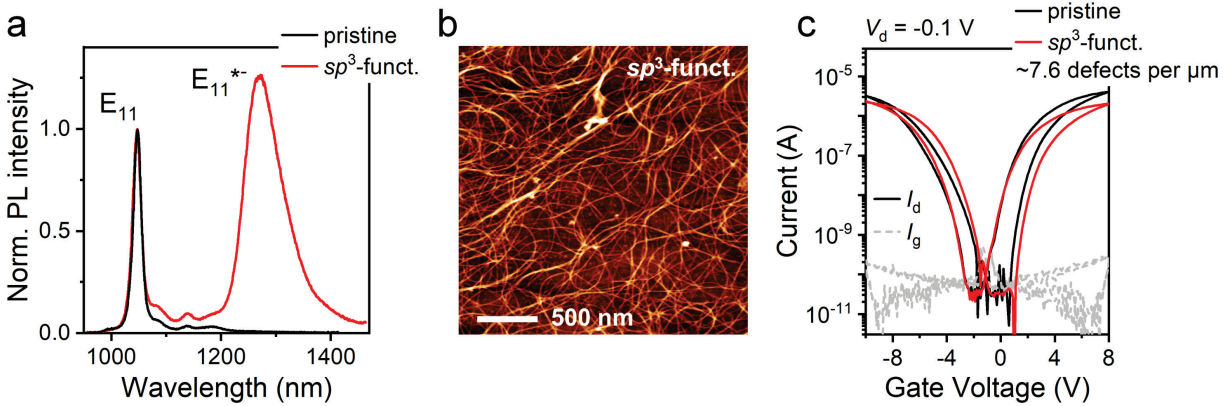

**Figure 3.** (a) Normalized PL spectra of a pristine (7,5) SWCNT dispersion and a dispersion after functionalization. Emission peaks corresponding to mobile $E_{11}$ excitons and defect-localized $E_{11}^{*-}$ excitons are labelled. (b) Representative atomic force micrograph of an aerosol-jet-printed network of $sp^3$-functionalized (7,5) SWCNTs with $E_{11}^{*-}$ defects. The scale bar is 500 nm. (c) Ambipolar transfer characteristics of pristine and functionalized (7,5) SWCNT network FETs in the linear regime ($V_d$ = -0.1 V). Solid lines are drain currents $I_d$, grey dashed lines represent gate leakage currents $I_g$.

Further insights into the possible trapping of charge carriers by sp³ defects can be gained from in-situ PL spectroscopy of the nanotube networks when electrostatically doped in FETs. For that, only a very small drain bias was applied (-0.01 V) to ensure a nearly uniform distribution of charges within the channel while still being able to record transfer characteristics. The gate voltage was increased stepwise to control the induced charge carrier density (holes or electrons). The nanotube network in the middle of the channel was excited with a 785 nm laser diode and PL spectra were recorded using an InGaAs line camera (see Experimental Section). Charge carriers in semiconducting SWCNTs lead to efficient PL quenching through Auger recombination as well as the formation of trions (charged excitons).[39-41] The overall reduced emission intensity from functionalized (6,5) nanotube networks with increasing concentration of holes (for negative applied gate voltages) is shown in Figure 4a. However, the normalized spectra (Figure 4b) clearly show stronger quenching of $E_{11}^{*-}$ defect emission compared to $E_{11}$ for the same carrier densities (i.e., gate voltages) as well as the emergence of trion emission at 1178 nm. Note that for in-situ PL measurements of (6,5) nanotubes with $E_{11}^*$ defects, the trion and defect emission peaks overlap, thus making an independent analysis difficult.[33] The data here also clearly show that trion and defect emission have different origins and are not related.

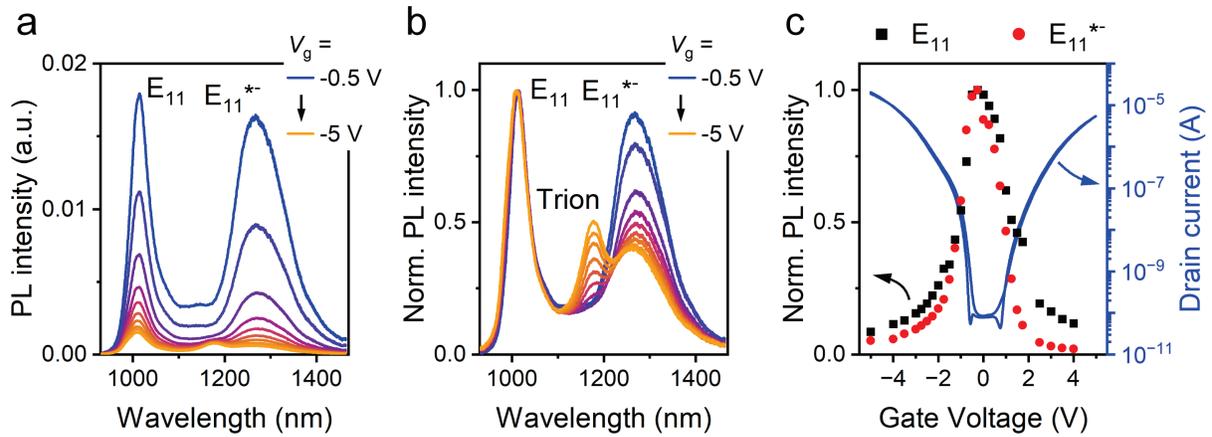

**Figure 4.** (a) Absolute and (b) normalized PL spectra (non-resonant excitation at 785 nm) of a (6,5) SWCNT network FET with $E_{11}^{*-}$ defects at different gate voltages $V_g$ in hole accumulation ($V_d$ = -0.01 V). At higher $V_g$, the trion emission becomes visible in the normalized spectra. (c) PL intensities of $E_{11}$ (black squares) and $E_{11}^{*-}$ (red circles) obtained from peak fits and normalized to their maxima. A corresponding transfer curve (linear regime, $V_d$ = -0.1 V) is shown in blue (right axis).

When plotting the normalized PL intensities of $E_{11}^{*-}$ and $E_{11}$ emission (resolved by fitting as Gaussian peaks) versus the applied gate voltage (Figure 4c), a small voltage offset in the drop of the PL intensity for both hole and electron accumulation becomes apparent. As introduced by Tanaka *et al.* for spectro-electrochemical measurements of pristine nanotubes[42] and later applied to functionalized SWCNTs, [43, 44] fitting the PL intensity versus gate voltage to a Nernst-type equation enables the extraction of voltage differences ($\Delta V_{g,1/2}$) between hole and electron accumulation that can be associated with the effective bandgap. For the $E_{11}$ emission, $\Delta V_{g,1/2}$ was determined to be 2.87 V versus 2.17 V for $E_{11}^{*-}$ (see Figure S3 and Table S2, Supporting Information), suggesting a narrower effective bandgap around the defect. Very similar PL quenching behavior was observed for functionalized (7,5) nanotubes as shown in Figure 5. The corresponding voltage differences $\Delta V_{g,1/2}$ were 2.42 V for quenching of $E_{11}$ emission and 1.86 V for $E_{11}^{*-}$ (see Figure S4 and Table S2, Supporting Information). As expected, the difference between $\Delta V_{g,1/2}$ for $E_{11}$ of (6,5) and (7,5) SWCNTs correlates with their electronic bandgaps (1.27 eV and 1.21 eV, respectively) but is not a quantitative value. The smaller offset between $\Delta V_{g,1/2}$ for $E_{11}$ and $E_{11}^{*-}$ of (6,5) and (7,5) SWCNTs (0.7 V versus 0.56 V) also scales roughly with the difference in the optical trap depths (268 meV versus 208 meV). These measurements suggest that $E_{11}^{*-}$ defects not only localize excitons but are also directly related to changes in the local valence and conduction band energies compared to the surrounding pristine nanotube and can thus act as shallow traps for both holes and electrons. Given the nearly identical reductions in hole and electron mobility it is reasonable to assume that the charge trap depths are very similar although electron-withdrawing or -pushing substituents may have an additional impact.[44, 45]

The gate voltage-dependent PL quenching indicated that charges could be trapped at the defect sites. To obtain more detailed information about the interaction of charge carriers with $E_{11}^{*-}$ defects and defect-localized excitons, PL lifetime measurements by TCSPC were performed on electrostatically doped (gated) nanotube networks. Note that analysis of the very short lifetime of the $E_{11}$ emission (*i.e.*, mobile excitons) in SWCNT networks (< 10 ps) was not possible due to limitation by the instrument response function (IRF) of the employed setup. However, in the undoped state, PL traces of the $E_{11}^{*-}$ emission could be fitted with a biexponential decay with a short (83 ps) and long (281 ps) lifetime component (see Figure S5, Supporting Information) as commonly observed for sp³ defects at room temperature.[46] The short lifetime might be assigned to the redistribution of the exciton population over the defect-state manifold, whereas the long lifetime encompasses all radiative and non-radiative decay pathways of defect-localized excitons.[47] Both lifetimes progressively decrease (to 32 ps and 129 ps, respectively) with increasing carrier density (here only shown for hole accumulation). These trends are in good agreement with charge-induced Auger-quenching as a very fast and efficient non-radiative decay path for defect-localized excitons. The PL lifetime data further corroborate the notion that charge carriers are indeed trapped at $E_{11}^{*-}$ defect sites, leading to more non-radiative exciton decay as well as lower carrier mobilities.

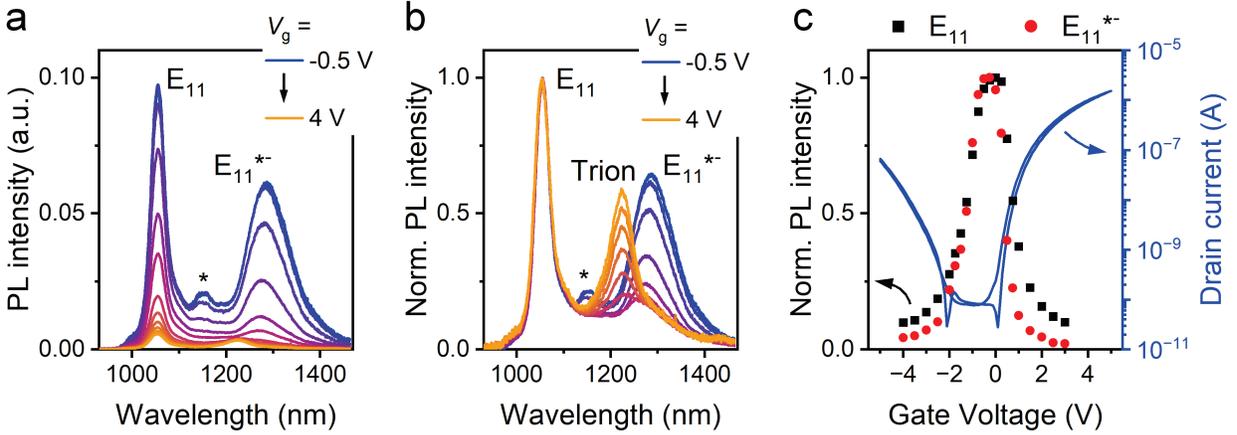

**Figure 5.** (a) Absolute and (b) normalized PL spectra (non-resonant excitation at 785 nm) of a (7,5) SWCNT network FET with $E_{11}^{*-}$ defects at different gate voltages $V_g$ in electron accumulation ($V_d$ = -0.01 V). At high voltages, trion emission becomes visible in the normalized spectra. The asterisk marks the $E_{11}$ emission peak of residual (7,6) SWCNTs. (c) PL intensities of $E_{11}$ (black squares) and $E_{11}^{*-}$ (red circles) obtained from peak fits and normalized to their maxima and corresponding transfer curve (linear regime, $V_d$ = -0.1 V) in blue (right axis).

Finally, the EL from (6,5) and (7,5) nanotubes with $E_{11}^{*-}$ defects was explored. Despite the indicated charge trapping at defects, the SWCNT networks still showed balanced hole and electron transport even at reasonably high defect densities (see above), which is required for light-emitting FETs. The respective FETs were operated in the ambipolar regime at high drain voltages to accumulate both holes and electrons in the channel and create a recombination and emission zone as described in detail elsewhere.[48] The applied voltages were chosen such that the emission zone (width 1-2 µm) was located in the middle of the channel (channel length 40 µm) to avoid any potential effects of carrier imbalance, contact resistance or the edge of the metal electrodes. The overall drain current (proportional to the number of hole and electron recombination events) was varied by adjusting the voltages and was kept constant throughout the collection of EL spectra. Figure 6a shows the obtained EL spectra from functionalized (6,5) SWCNTs with clear $E_{11}$ emission at ~1015 nm as well as emission from the $E_{11}^{*-}$ defects at ~1280 nm. Importantly, the maximum of the defect emission peak was located well within the telecommunication O-band (1260-1360 nm), and at low drain currents more than 85% of photons were emitted through this defect channel. The overall external quantum efficiency for these devices was estimated to be 0.015% and thus similar to previous light-emitting FETs with functionalized SWCNTs.[33]

As the drain current was increased the integrated EL intensity also increased as shown for light-emitting FETs with functionalized (6,5) SWCNTs (Figure 6a) and for pristine (6,5) nanotubes in Figure S6 (Supporting Information). Normalization of the spectra to the $E_{11}$ emission peak (Figure 6b) demonstrates that $E_{11}^{*-}$ defect emission dominates at low current densities but saturates compared to the $E_{11}$ emission at higher current densities. This trend is also reflected in the double logarithmic plot of integrated EL peak intensities versus drain current (Figure 6c). The slope for defect emission is slightly lower than for mobile exciton emission, although both are close to unity and hence close to a linear increase of emission with current density.

These different dependencies are not due to trapping of excitons at the defect sites and subsequent state filling as observed for pulsed excitation of functionalized nanotubes at high excitation power.[25, 28] The estimated exciton density at the given current densities is still too low for this effect. It is more likely that trapping of holes and electrons at the defects leads to more efficient Auger quenching of defect emission at higher carrier densities similar to the stronger PL quenching discussed above. Note that such strong current density dependence of defect emission was not observed for the $E_{11}^{*}$ defect emission of (6,5) SWCNTs,[33] probably due to the lower trap depth (160-190 meV) compared to $E_{11}^{*-}$ defects (268 meV).

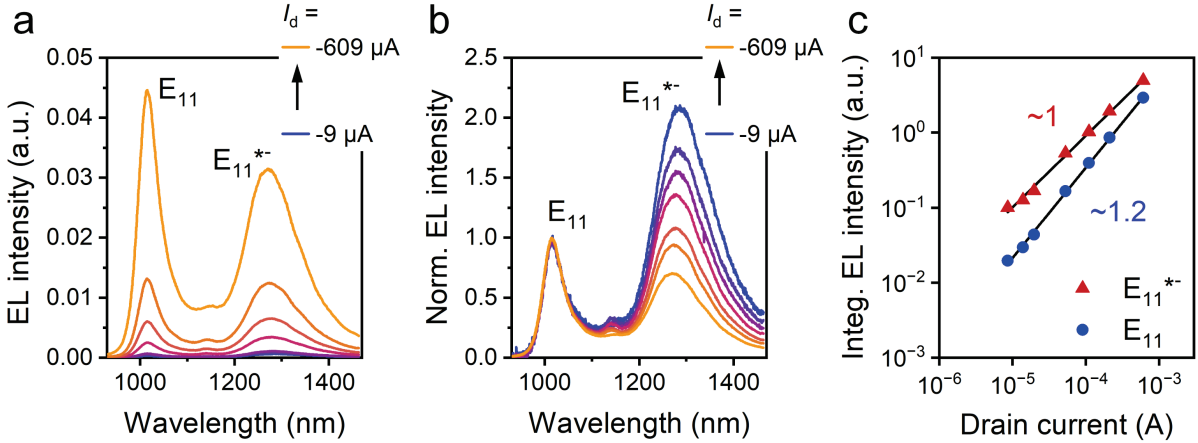

**Figure 6.** (a) Absolute and (b) normalized EL spectra of a (6,5) SWCNT network light-emitting FETs with $E_{11}^{*-}$ defects in the ambipolar regime for different drain currents $I_d$. (c) Double-logarithmic plot of the integrated EL intensity versus $I_d$. Blue circles correspond to $E_{11}$ emission, red triangles denote $E_{11}^{*-}$ defect emission (integration from 950-1100 nm and 1150-1450 nm, respectively). Black solid lines are linear fits to the data, and numbers indicate the corresponding slopes.

The EL spectra from ambipolar light-emitting FETs based on functionalized (7,5) nanotubes are shown in Figure 7a. In addition to the emission peaks from mobile excitons of the (7,5) nanotubes at 1055 nm and residual (7,6) SWCNTs at 1150 nm, a significant number of photons originates from the $E_{11}^{*-}$ defects with emission centred around 1300 nm and thus fully within the O-band for telecommunication. The EL spectra of pristine (7,5) nanotubes only show $E_{11}$ emission from mobile excitons as well as some trion emission (~ 1225 nm) at higher current densities (see Figure S7, Supporting Information). The ratio of $E_{11}$ to $E_{11}^{*-}$ emission from light-emitting FETs with functionalized (7,5) SWCNTs (see normalized EL spectra in Figure 7b and double logarithmic plots in Figure 7c) does not change and the overall intensity increases roughly linearly with current. The reduced current density dependence of the $E_{11}^{*-}$ emission might be due to the limited drain current range (due to the lower carrier mobilities) and the lower trap depth compared to FETs with functionalized (6,5) nanotubes. Note that the low drain currents also prevented a reliable determination of external quantum efficiencies for EL from pristine or functionalized (7,5) nanotubes.

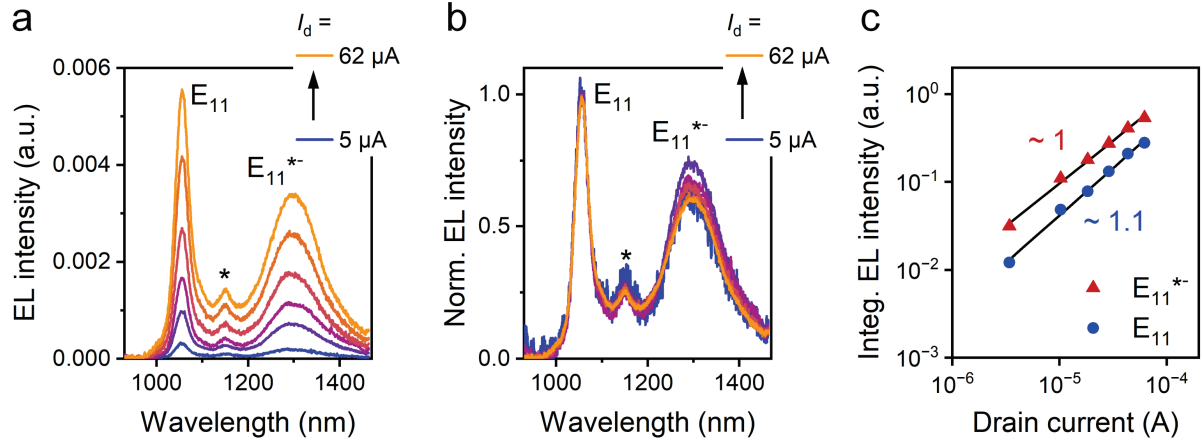

**Figure 7.** (a) Absolute and (b) normalized EL spectra of a pristine (7,5) SWCNT network light-emitting FET and a (7,5) SWCNT network FET with $E_{11}^{*-}$ defects in the ambipolar regime for different drain currents $I_d$. The asterisk marks the $E_{11}$ peak of residual (7,6) SWCNTs. (c) Double-logarithmic plot of the integrated EL intensity versus $I_d$. Blue circles correspond to $E_{11}$ emission, red triangles denote $E_{11}^{*-}$ defect emission (integration from 950-1120 nm and 1190-1450 nm, respectively). Black solid lines are linear fits to the data, and numbers indicate the corresponding slopes.

**Conclusions**

In summary, we have demonstrated near-infrared photoluminescence and electroluminescence in the telecommunication O-band from dense networks of $sp^3$-functionalized (6,5) and (7,5) single-walled carbon nanotubes in light-emitting FETs. The majority of photons are emitted from the luminescent defects at 1280 to 1300 nm and emission intensities scale roughly linearly with current density. The controlled introduction of $E_{11}^{*-}$ defects with further red-shifted emission, provides direct access to emission wavelengths beyond 1200 nm even with small-diameter nanotubes, which are easily purified and functionalized in large amounts. Furthermore, the investigated ambipolar light-emitting field-effect transistors gave insights into the impact of the specific luminescent defects and their optical trap depth on the local valence and conduction band energies around the defect site. Their charge trapping properties depend both on their binding configuration ($E_{11}^*$ or $E_{11}^{*-}$) and the diameter of the nanotube. The presented data showcase the spectral tunability of *sp*$^3$ defect emission as achieved by choice of the SWCNT species and control over the defect binding configuration through functionalization chemistry while retaining good charge transport. This combination should enable the application of covalently functionalized SWCNTs in near-infrared light-emitting devices, although improvements of the emission efficiency will be required.

**Experimental Section**

**Preparation of (6,5) and (7,5) SWCNT dispersions.**

Highly purified (6,5) and (7,5) SWCNT dispersions in toluene were obtained from CoMoCAT raw material (Sigma Aldrich, batch MKCJ7287, 0.4 g L$^{-1}$) *via* selective polymer-wrapping under shear force mixing (Silverson L2/Air mixer, 10230 rpm, 72 h, 20 °C) as described previously.[17] (6,5) SWCNTs were extracted with the fluorene-bipyridine copolymer poly[(9,9-dioctylfluorenyl-2,7-diyl)-*alt*-(6,6'-(2,2'-bipyridine))] (PFO-BPy, American Dye Source, $M_W$ = 40 kg mol$^{-1}$, 0.5 g L$^{-1}$), whereas (7,5) SWCNTs were selectively dispersed with poly(9,9-dioctylfluorene) (PFO, Sigma Aldrich, $M_W$ > 20 kg mol$^{-1}$, 0.9 g L$^{-1}$). The resulting dispersions after shear force mixing were centrifuged twice for 45 min at 60000*g* (Beckman Coulter Avanti J26XP centrifuge), and the combined supernatants were filtered through a polytetrafluoroethylene (PTFE) syringe filter (pore size 5 μm) to remove larger aggregates.

**Covalent functionalization of polymer-wrapped (6,5) and (7,5) SWCNTs**

Luminescent E$_{11}$*$^-$ defects were introduced to polymer-wrapped (6,5) and (7,5) SWCNTs according to the protocol developed by Settele *et al*.[25] To reduce the content of unbound wrapping polymer prior to the functionalization reaction, the dispersions obtained by shear force mixing were filtered through a PTFE membrane filter (Merck Millipore JVWP, 0.1 μm pore size), washed with toluene (10 mL), and re-dispersed in fresh toluene by bath sonication. SWCNTs were functionalized with 2-iodoaniline (Sigma Aldrich, 98%) in the presence of the strong organic base potassium *tert*-butoxide (KO*t*Bu, Sigma Aldrich, 98%). First, the aniline compound was dissolved in toluene, and dimethyl sulfoxide (DMSO, Sigma Aldrich, anhydrous) and a solution of KO*t*Bu in tetrahydrofuran (THF, Sigma Aldrich, anhydrous) were added to the solution. Then, the polymer-depleted SWCNT dispersion in toluene was added to the reaction mixture so that the optical density at the $E_{11}$ absorption peak was 0.3 cm$^{-1}$ (concentrations of 0.54 µg mL$^{-1}$ for (6,5) SWCNTs and 0.86 µg mL$^{-1}$ for (7,5) SWCNTs). The concentrations of the aniline compound and KO*t*Bu in the final reaction mixture were 29.3 mmol L$^{-1}$ and 58.6 mmol L$^{-1}$, respectively, and the volumetric fractions were 83.3:8.3:8.3 toluene:DMSO:THF. The reaction proceeded under stirring and in the dark over 30-180 min before it was terminated by vacuum filtration through a PTFE membrane filter (Merck Millipore JVWP, 0.1 µm pore size). The collected nanotubes were thoroughly washed with methanol and toluene (10 mL each) to remove unreacted compounds and by-products.

**Fabrication of SWCNT network field-effect transistors**

FETs were fabricated on glass substrates (Schott AF32eco, 300 µm thickness) that were passivated with the cross-linked polymer divinyltetramethylsiloxane-bis-benzocyclobutene (BCB).[35] Glass substrates were cleaned by ultrasonication in acetone and 2-propanol (10 min each). In a nitrogen-filled glovebox, a resin of BCB (Cyclotene 3022-35) was diluted with mesitylene, spin-coated onto the substrates (500 rpm for 3 s, then 8000 rpm for 60 s), and cross-linked *via* thermal annealing at 290 °C for 2 min. On these substrates, interdigitated bottom-contact electrodes (channel length $L$ = 20 µm, channel width $W$ = 10 mm) were patterned by photolithography (double-layer LOR5B/S1813 photoresist) and electron beam evaporation of chromium (3 nm) and gold (30 nm). After lift-off in *N*-methyl-2-pyrrolidone (NMP) over night and rinsing with acetone and 2-propanol, SWCNT networks were deposited onto the substrates as detailed below. All SWCNT networks were annealed in inert atmosphere at 150 °C for 45 min. Then, a double-layer gate dielectric was deposited, which consisted of a ~ 10-14 nm poly(methyl methacrylate) (PMMA) layer and ~ 60-65 nm hafnium oxide (HfO$_x$). First, a solution of PMMA (Polymer Source, $M_W$ = 315 kg mol$^{-1}$, syndiotactic) in *n*-butylacetate (6 g L$^{-1}$) was spin-coated at 4000 rpm for 60 s, followed by annealing at 80 °C for 2 min. Then, atomic layer deposition (Ultratech Savannah S100, 500 cycles, 100 °C) was used to deposit the HfO$_x$ layer of from a tetrakis(dimethylamino)hafnium precursor (Strem Chemicals) and water as the oxidizing agent. To complete the devices, silver top-gate electrodes (30 nm) were thermally evaporated through a shadow mask.

**Deposition of SWCNT networks.**

Concentrated dispersions of pristine and *sp*$^3$-functionalized (6,5) SWCNTs were obtained by bath sonication of SWCNT filter cakes in a small volume of fresh toluene with 1,10-phenanthroline (2.8 mmol L$^{-1}$) as stabilizer.[49] From such dispersions with an optical density of 8 cm$^{-1}$ at the $E_{11}$ absorption transition, SWCNT networks were created by repeated spin-coating (2000 rpm, 30 s, 3 times) onto the prepatterned bottom-contact electrodes with intermediate annealing steps (100 °C, 2 min). Subsequently, substrates were rinsed with THF and 2-propanol. To remove all nanotubes outside the channel area, photolithography was performed as described above, followed by oxygen plasma treatment and lift-off in NMP for 60 min. Pristine and *sp*$^3$-functionalized (7,5) SWCNT networks were deposited *via* aerosol-jet printing (Optomec AJ200 printer, 200 μm inner diameter nozzle).[50] Terpineol (Sigma Aldrich, mixture of isomers) was added (2 vol-%) to SWCNT dispersions in toluene (optical density of 1-2 cm$^{-1}$ at the $E_{11}$ absorption transition) obtained by bath sonication of filter cakes to increase the viscosity and improve aerosol formation. The ink was printed exclusively in the channel area of the devices, which removed the need for patterning of the networks. During the printing process, the ink was constantly sonicated, and the substrates were placed on a heated stage at 100 °C. Finally, the substrates were rinsed with THF and 2-propanol to remove residual polymer and terpineol.

**Spectroscopic characterization**

PL spectra of SWCNT dispersions and thin films were measured with a home-built setup. Samples were excited either with the wavelength-filtered output of a picosecond-pulsed supercontinuum laser (NKT Photonics SuperK Extreme, 19.5 MHz repetition rate) or with a 785 nm laser diode (Alphalas Picopower-LD-785-50) in continuous wave mode. The laser beam was focused on the sample (SWCNT dispersion in a quartz glass cuvette or SWCNT network FETs on transparent glass substrates) with a NIR-optimized 50× objective (Olympus, N.A. 0.65). Emitted photons were collected with the same objective, and scattered laser light was blocked by appropriate long-pass filters. The emission was spectrally resolved with a grating spectrograph (Acton SpectraPro SP2358, grating blaze 1200 nm, 150 lines mm$^{-1}$) and detected with a liquid nitrogen-cooled InGaAs line camera (Princeton Instruments OMA V:1024-1.7 LN). For time-resolved PL measurements in a time-correlated single-photon counting scheme, the spectrally selected emission was focused onto a gated InGaAs/InP avalanche photodiode (Micro Photon Devices) with a 20× objective (Mitutoyo, N.A. 0.40). Histograms of photon arrival times were recorded with a counting module (PicoQuant PicoHarp 300) and fitted with biexponential functions in a reconvolution procedure using the SymPhoTime64 software. The instrument-limited decay of $E_{11}$ excitons in a dense (6,5) SWCNT film served as the instrument response function. For EL spectroscopy and PL spectroscopy at a constant gate voltage, SWCNT network FETs were electrically contacted, and voltages were applied with a Keithley 2612A source meter. EL spectra were always recorded from the center of the channel to avoid possible effects of contact resistance or impact excitation at the electrode edges. All spectra were corrected to account for the absorption of optics in the detection path and the wavelength-dependent detector efficiency.

**Additional characterization techniques**

Baseline-corrected absorption spectra of SWCNT dispersions were acquired with a Varian Cary6000i UV-vis-NIR spectrometer using quartz glass cuvettes with 1 cm path length. Raman spectra (532 nm and 633 nm excitation) were measured with a Renishaw inVia Reflex confocal Raman microscope with a 50× objective (N.A. 0.50) in backscattering configuration. Typically, Raman data were obtained by averaging over 2500 individual spectra collected from an area of 100×100 µm$^2$. Atomic force micrographs of SWCNT networks were recorded with a Bruker Dimension Icon atomic force microscope in ScanAsyst mode. Thicknesses of polymer and HfO$_x$ films were determined with a Bruker DektakXT Stylus profilometer. Current-voltage characteristics of SWCNT network FETs were recorded with an Agilent 4156C semiconductor parameter analyzer. Effective device capacitances were extracted from capacitance versus gate voltage sweeps with a Solatron Analytical ModuLab XM MTS impedance analyzer at a frequency of 100 Hz. They varied between 130-137 nF cm$^{-2}$ for (6,5) SWCNT network FETs and 87-88 nF cm$^{-2}$ for (7,5) SWCNT network FETs.

## AUTHOR CONTRIBUTIONS

N.F.Z. fabricated and measured all samples and analyzed the data. S.S. and F.L.S. fabricated selected samples, contributed to characterization and prepared chemically functionalized SWCNT dispersions. S.L. prepared (7,5) SWCNT dispersions. J.Z. conceived and supervised the project. N.F.Z. and J.Z. wrote the manuscript with input from all authors.

## ACKNOWLEDGMENT

This project has received funding from the European Research Council (ERC) under the European Union's Horizon 2020 research and innovation programme (Grant Agreement No. 817494 "TRIFECTs").
23
## AUTHOR CONTRIBUTIONS

N.F.Z. fabricated and measured all samples and analyzed the data. S.S. and F.L.S. fabricated selected samples, contributed to characterization and prepared chemically functionalized SWCNT dispersions. S.L. prepared (7,5) SWCNT dispersions. J.Z. conceived and supervised the project. N.F.Z. and J.Z. wrote the manuscript with input from all authors.

## ACKNOWLEDGMENT

This project has received funding from the European Research Council (ERC) under the European Union's Horizon 2020 research and innovation programme (Grant Agreement No. 817494 "TRIFECTs").

# Supporting Information

Tuning Electroluminescence from Functionalized SWCNT Networks further into the Near-Infrared

*Nicolas F. Zorn, Simon Settele, Finn L. Sebastian, Sebastian Lindenthal and Jana Zaumseil\**

Institute for Physical Chemistry, Universität Heidelberg, D-69120 Heidelberg, Germany

**Corresponding Author:** *zaumseil@uni-heidelberg.de

# Contents

# Material Characterization

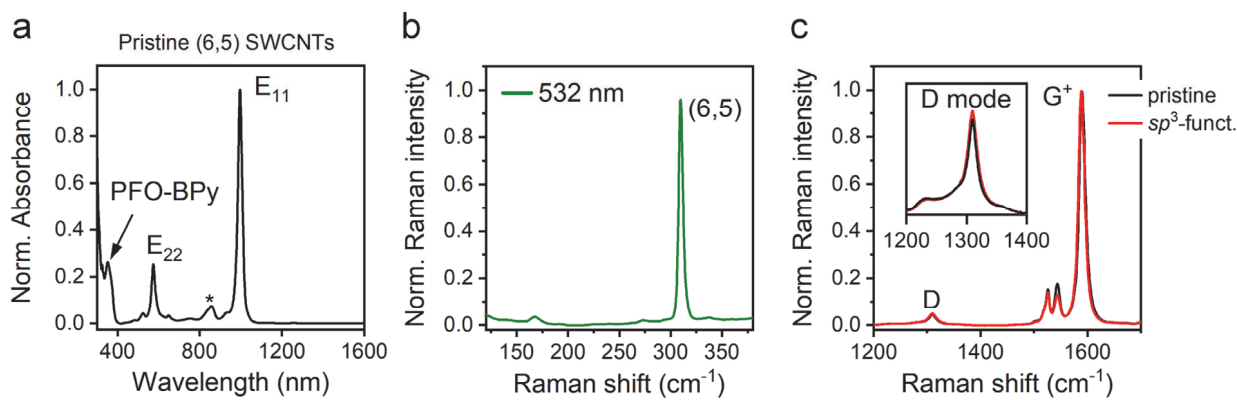

**Figure S1.** (a) Absorption spectrum of a (6,5) SWCNT dispersion with PFO-BPy as wrapping polymer in toluene. The main excitonic transitions of (6,5) SWCNTs ($E_{11}$, $E_{22}$) and the polymer absorption band are labelled. The asterisk marks the $E_{11}$ phonon sideband of (6,5) SWCNTs. (b) Radial breathing mode (RBM) range of normalized Raman spectrum (532 nm excitation) of purified (6,5) SWCNTs. (c) Resonant Raman spectra (532 nm excitation) of pristine and functionalized (6,5) SWCNTs normalized to the $G^+$ mode. The inset shows a zoom-in on the defect-related D mode.

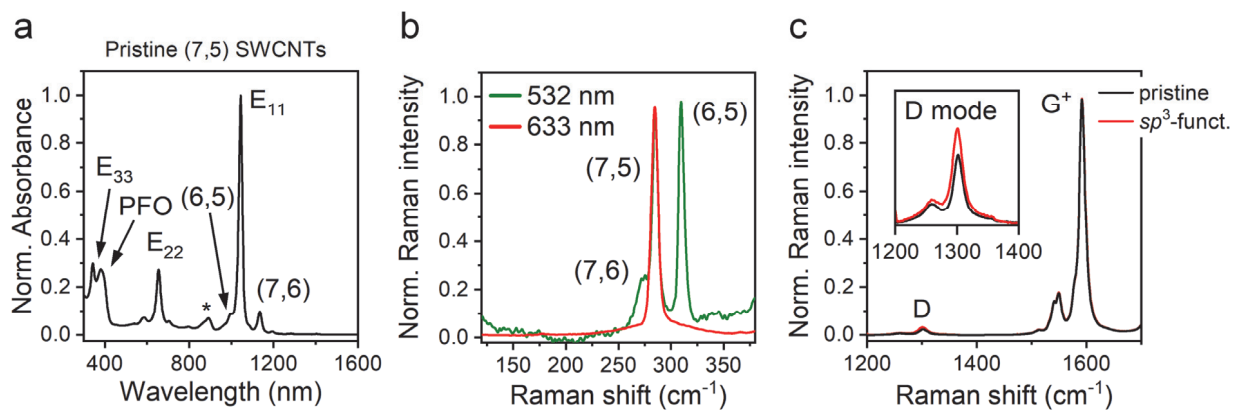

**Figure S2.** (a) Absorption spectrum of a (7,5) SWCNT dispersion with PFO as the wrapping polymer in toluene. The main excitonic transitions of (7,5) SWCNTs ($E_{11}$, $E_{22}$, $E_{33}$), the polymer absorption band, and peaks corresponding to the $E_{11}$ transitions of residual (6,5) and (7,6) SWCNTs are labelled. The asterisk marks the $E_{11}$ phonon sideband of (7,5) SWCNTs. (b) Radial breathing mode (RBM) range of normalized Raman spectra (532 nm and 633 nm excitation) of purified (7,5) SWCNTs also showing residual signals for (6,5) and (7,6) SWCNTs. (c) Resonant Raman spectra (633 nm excitation) of pristine and functionalized (7,5) SWCNTs normalized to the $G^+$ mode. The inset shows a zoom-in on the defect-related D mode.

**Table S1.** Calculated defect density (from change of integrated D/G$^+$ ratio, see Raman spectra above) of functionalized (6,5) and (7,5) SWCNTs and corresponding hole and electron mobilities of network field-effect transistors (FETs) compared to pristine SWCNT network FETs (as shown in Figures 2c and 3c of the main text).

| Sample | Defect Density (µm$^{-1}$) | Hole Mobility (cm²V$^{-1}$s$^{-1}$) | Electron Mobility (cm²V$^{-1}$s$^{-1}$) |
|---|---|---|---|
| (6,5) SWCNTs | | | |
| Pristine | - | 3.9 ± 0.3 | 2.1 ± 0.3 |
| Functionalized | 3.5 | 2.2 ± 0.1 | 1.1 ± 0.1 |
| (7,5) SWCNTs | | | |
| Pristine | - | 0.14 ± 0.01 | 0.14 ± 0.01 |
| Functionalized | 7.6 | 0.07 ± 0.01 | 0.06 ± 0.01 |

## Gate Voltage-Dependent PL Quenching

The gate voltage-dependent photoluminescence quenching for hole and electron accumulation was fitted with Nernst-type equations as introduced by Shiraishi et al.[1] with

$$\frac{\mathrm{PL}(V_g)}{\mathrm{PL}_{max}} = \frac{1}{1+\exp\left(A\cdot\left(V_{g,1/2}^h - V_g\right)\right)} \quad (1)$$

for hole accumulation and

$$\frac{\mathrm{PL}(V_g)}{\mathrm{PL}_{max}} = \frac{1}{1+\exp\left(A\cdot\left(V_g - V_{g,1/2}^e\right)\right)} \quad (2)$$

for electron accumulation. Finally, the value $\Delta V_{g,1/2} = V_{g,1/2}^e - V_{g,1/2}^h$ was extracted, similar to an electrochemical bandgap (see values in **Table S2**).

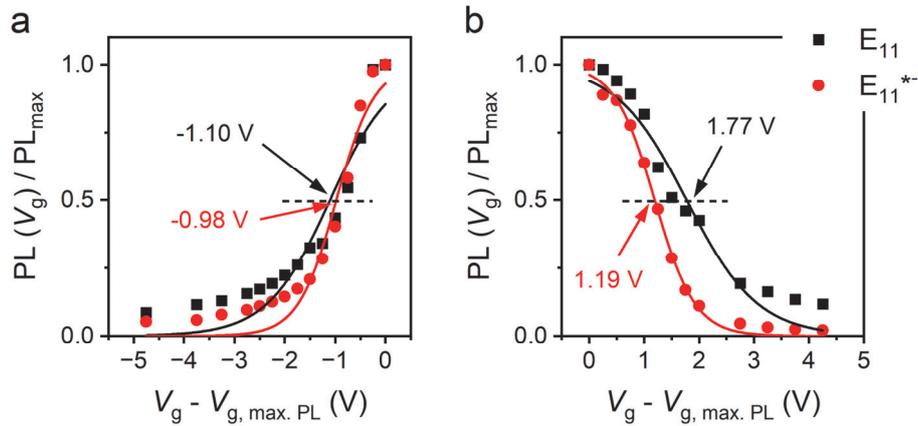

**Figure S3.** PL intensities of $E_{11}$ (black squares) and $E_{11}^{*-}$ (red circles) for (a) hole accumulation and (b) electron accumulation obtained from fits to the PL spectra (non-resonant excitation at 785 nm) of a (6,5) SWCNT network light-emitting FET with $E_{11}^{*-}$ defects and normalized to their maxima. Note that the x-axis was adjusted in order to shift the PL maximum (i.e., undoped state) to zero voltage. Solid lines are fits according to equations (1) and (2). Gate voltages corresponding to half the initial PL intensity are indicated.

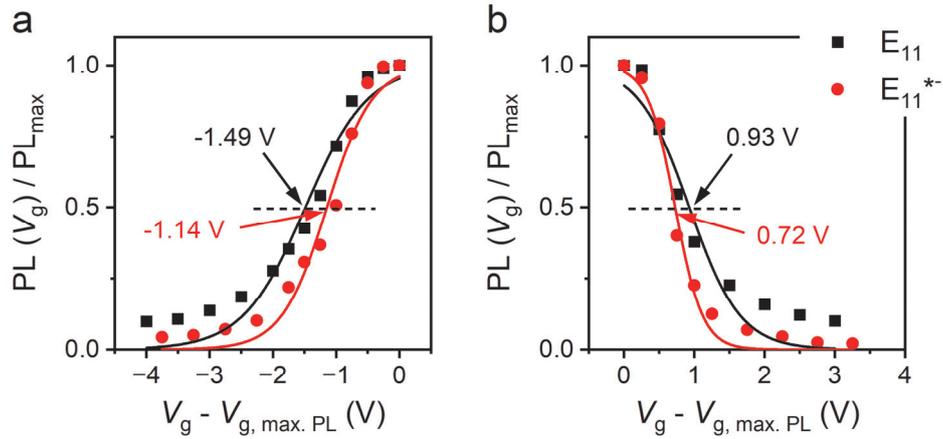

**Figure S4.** PL intensities of $E_{11}$ (black squares) and $E_{11}^{*-}$ (red circles) in (a) hole accumulation and (b) electron accumulation obtained from fits to the PL spectra (non-resonant excitation at 785 nm) of a (7,5) SWCNT network light-emitting FET with $E_{11}^{*-}$ defects and normalized to their maxima. Note that the x-axis was adjusted in order to shift the PL maximum (i.e., undoped state) to zero voltage. Solid lines are fits according to equations (1) and (2). Gate voltages corresponding to half the initial PL intensity are indicated.

**Table S2.** Gate voltage-dependent PL quenching of $E_{11}$ and $E_{11}^{*-}$ in (6,5) and (7,5) SWCNT network light-emitting FETs. Voltages ($V_{g,1/2}^{h,e}$) corresponding to half the initial PL intensity in hole and electron accumulation and voltage differences ($\Delta V_{g,1/2}$) were obtained from fits to equations (1) and (2).

| Nanotube species | State | $V_{g,1/2}^{h}$ (V) | $V_{g,1/2}^{e}$ (V) | $\Delta V_{g,1/2}$ (V) | Difference between $E_{11}$ and $E_{11}^{*-}$ |
|---|---|---|---|---|---|
| (6,5) | $E_{11}$ | -1.10 | 1.77 | 2.87 | 0.7 V |
| | $E_{11}^{*-}$ | -0.98 | 1.19 | 2.17 | |
| (7,5) | $E_{11}$ | -1.49 | 0.93 | 2.42 | 0.56 V |
| | $E_{11}^{*-}$ | -1.14 | 0.72 | 1.86 | |

## Time-Resolved PL Spectroscopy

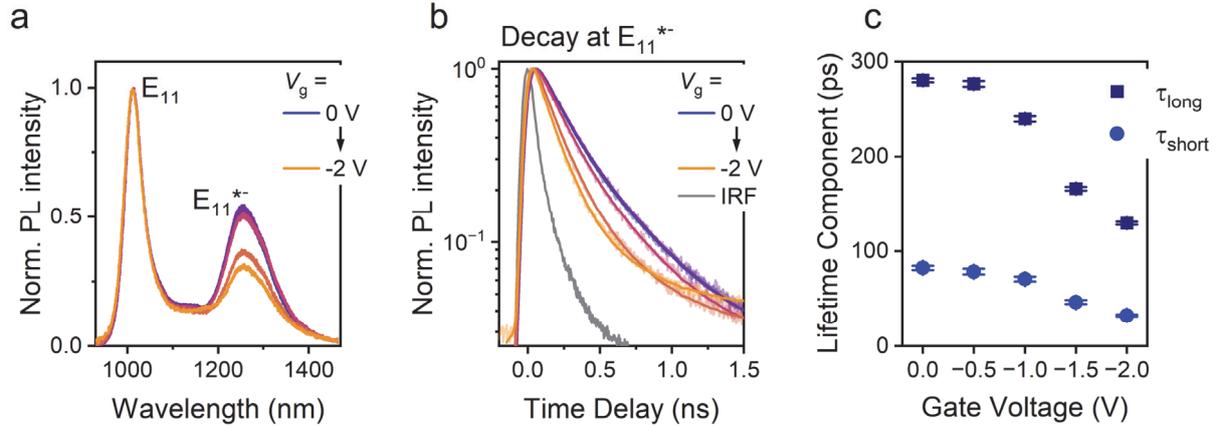

**Figure S5.** (a) Normalized PL spectra (pulsed $E_{22}$ excitation at 575 nm) of a (6,5) SWCNT network light-emitting FET with $E_{11}*^-$ defects at different gate voltages $V_g$ (voltage steps, 0.5 V) in hole accumulation ($V_d$ = -0.01 V). (b) Time-resolved PL decay traces at the $E_{11}*^-$ defect emission wavelength (1260 nm) at different $V_g$ corresponding to the spectra in (a). Measured data are shown in light colors, biexponential fits to the data are shown in darker colors. The instrument response function (IRF) is shown in grey. (c) Extracted lifetime components ($\tau_{long}$, $\tau_{short}$) obtained from biexponential fits in (b) versus applied gate voltage.

**Table S3.** Gate voltage-dependent $E_{11}*^-$ PL decay in an sp³-functionalized (6,5) SWCNT network light-emitting FET. Long ($\tau_{long}$) and short ($\tau_{short}$) lifetime components, normalized amplitudes ($A_{long}$, $A_{short}$), and amplitude-averaged lifetimes ($\tau_{ampl.-avg.}$) were obtained from biexponential fits to the PL decays at the $E_{11}*^-$ emission wavelength (1260 nm).

| Gate Voltage (V) | $\tau_{long}$ (ps) | $A_{short}$ | $\tau_{short}$ (ps) | $A_{short}$ | $\tau_{ampl.-avg.}$ (ps) |
|---|---|---|---|---|---|
| 0 | 281 | 0.35 | 83 | 0.65 | 152 |
| -0.5 | 277 | 0.38 | 79 | 0.62 | 153 |
| -1.0 | 239 | 0.36 | 71 | 0.64 | 132 |
| -1.5 | 166 | 0.32 | 47 | 0.68 | 84 |
| -2.0 | 129 | 0.29 | 32 | 0.71 | 61 |

## Electroluminescence from Pristine SWCNT Networks

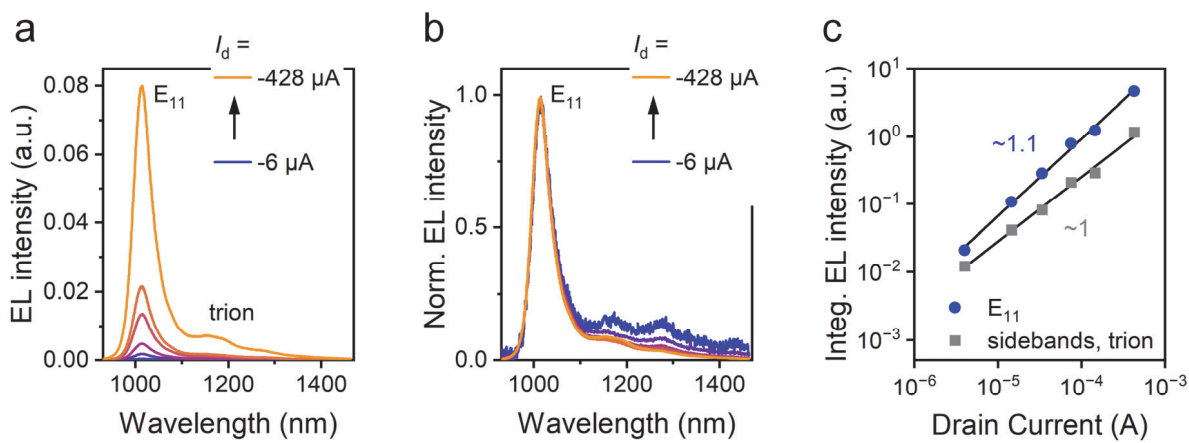

**Figure S6.** (a) Absolute and (b) normalized EL spectra of a pristine (6,5) SWCNT network light-emitting FET in the ambipolar regime for different drain currents $I_d$. (c) Double-logarithmic plot of the integrated EL intensity versus drain current. Blue circles are $E_{11}$ emission intensities, grey squares correspond to red-shifted sidebands and trion contributions (integration from 950-1100 nm and 1100-1400 nm, respectively). Black solid lines are linear fits to the data, and numbers indicate the corresponding slopes.

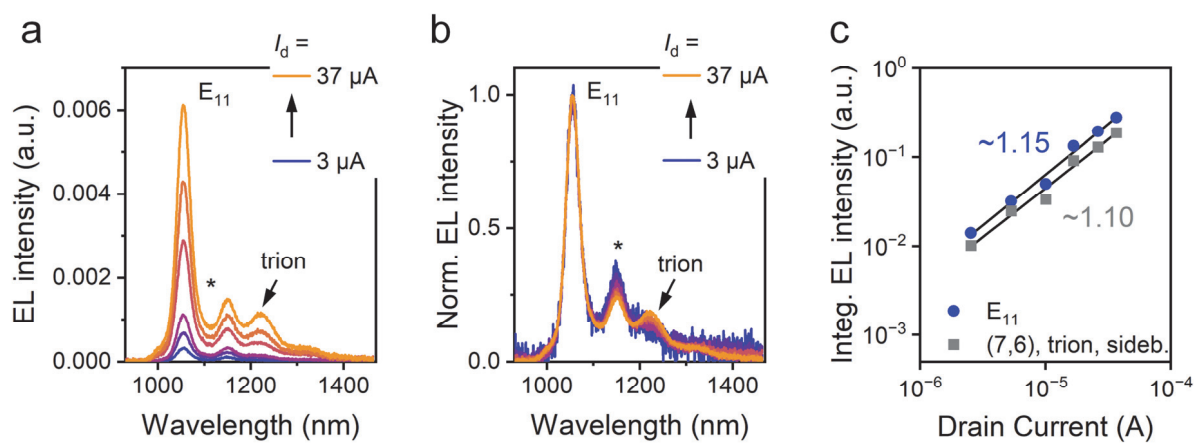

**Figure S7.** (a) Absolute and (b) normalized EL spectra of a pristine (7,5) SWCNT network light-emitting FET in the ambipolar regime for different drain currents $I_d$. (c) Double-logarithmic plot of the integrated EL intensities versus drain current. Blue circles are $E_{11}$ emission intensities, grey squares correspond to emission from (7,6) SWCNTs (marked by asterisk), red-shifted sidebands and trion contributions (integration from 950-1120 nm and 1120-1400 nm, respectively). Black solid lines are linear fits to the data, and numbers indicate the corresponding slopes.